# Spontaneous Nanopatterning and Strain Relaxation in SiGe Layers Grown by Oxidative Solid Phase Epitaxy

*Sophie E. Bierer, Trevor R. Smith, Arezoo Mafi, Sunzhuoran Wang, Chengqian Liao, Vatsalkumar Patel, Andrew P. Knights, Ryan B. Lewis**

S.E. Bierer, T.R. Smith, A. Mafi, S. Wang, C. Liao, V. Patel, A.P. Knights, R.B. Lewis
Engineering Physics, McMaster University, Hamilton, Canada
E-mail: rlewis@mcmaster.ca

Funding:
Financial support is from the Natural Sciences and Engineering Research Council of Canada under projects RGPIN-2020-05721 and ALLRP-588298-23.

**Abstract:**
The wafer-scale monolithic integration of III-V materials on Si would lead to revolutionary optoelectronic hardware for data, computing and other applications. However, heteroepitaxy of III-Vs on Si requires overcoming the large lattice and thermal mismatches between the materials and reducing threading dislocations densities. In this work, we explore the oxidative solid phase epitaxy (SPE) of $Ge^+$ implanted Si(111) to form ultra-thin strain-relieving SiGe metamorphic buffer layers for heteroepitaxy on Si. The SPE process is shown to result in a nanopatterning of the Ge concentration variation across the sample surface, visible by scanning and transmission electron microscopy (SEM and TEM). The concentration patterning is the result of a hexagonal network of Shockley partial dislocations at the SiGe/Si interface. Analyzing the pattern spacing observed by SEM is demonstrated as an easy, non-destructive method for obtaining the local strain state of SiGe layers. This work is important for engineering ultra-thin SiGe metamorphic buffer layers for III-V optoelectronics heteroepitaxy on the silicon platform.

## 1. Introduction

The integration of III-V materials onto Si enables efficient, high-speed III-V optoelectronic devices to be combined with Si platforms for electronics and photonic integrated circuits.[1–3] Achieving wafer-scale epitaxial growth of III-Vs on Si, however, requires overcoming issues related to lattice parameter, polarity and thermal mismatches. If not mitigated, these issues lead to high threading dislocation densities and other defects, degrading device performance.[4–6] One solution is to incorporate dislocation-filtering metamorphic buffer layers between Si and the III-V.[7–9] A popular buffer material is Ge, as its mismatch from GaAs is only 0.08%. Dislocation-filtering Ge buffer layers are typically grown several microns thick to reduce the density of threading dislocations in the subsequently grown III-V layers.[10–13] Thick buffer layers, however, have drawbacks due to the thermal expansion mismatch, and pose challenges for optical coupling between components.

While most SiGe/Si growth studies have been carried out on Si(100), the (111) orientation has been shown to exhibit interesting dislocation formation mechanisms that could be beneficial for III-V integration. Specifically, for SiGe growth on Si(111), it has been shown that using an Sb surfactant induces two-dimensional layer growth over the usual island formation.[14–16] In these layers, Shockley partial dislocations were shown to form in pairs that self-annihilate their own threading defects. The result is a SiGe epilayer that contains minimal threading defects and is highly efficient at relieving strain by interface misfit dislocations.

One method for creating thin SiGe layers on Si involves forming a SiGe layer (e.g. through $Ge^+$ ion implantation or SiGe deposition), followed by an oxidation process precipitating/condensing a crystalline Ge-rich SiGe layer at the interface between Si and the oxide.[17–19] This process has recently been employed on Si(111) to create single crystal sub-20-nm-thick SiGe buffer layers, on which GaAs epitaxial growth was demonstrated.[20] This constitutes a novel approach for III-V heteroepitaxy on Si without the need for thick dislocation filtering buffer layers. The details of the strain-relieving dislocations and the strain state of these solid-phase-epitaxy-grown ultra-thin SiGe layers, however, remain unexplored.

In this work, we examine the microstructure and composition of SiGe/Si(111) heterostructures formed by the oxidative solid phase epitaxy (SPE) of Si(111) wafers implanted with varying concentrations of Ge. The SiGe/Si interface is found to contain a hexagonal network of strain relieving dislocations that run along $\langle\bar{1}10\rangle$ directions. This periodic dislocation array results in nano-scale patterning of the SiGe composition, which is analyzed by scanning electron microscopy (SEM) and plan-view transmission electron microscopy (TEM). The SEM pattern provides a simple, non-destructive means of determining the local strain state of the SiGe layer. Finally, the periodicity of the composition fluctuations seen in SEM are analyzed with a pattern recognition algorithm, yielding the local strain state of the SiGe films. These results provide important information on the strain and composition of ultra-thin SiGe metamorphic layers on Si(111), required to further develop this platform for III-V-on-Si integration.

## 2. Results and Discussion

A series of SiGe/Si(111) samples with initial $Ge^+$ ion doses of 1.50, 2.25, and 3.00 × $10^{16}$ $cm^{-2}$ were characterized (samples A–C, respectively). **Figures 1a–c** show normalized Auger electron spectroscopy (AES) depth profiles of Si and Ge for samples A–C. In the concentration profiles, a depth of zero corresponds to the top of the SiGe layer. All samples show a Ge concentration maximum within 1 nm of top of the SiGe layer, followed by a gradual decrease to zero with increasing depth. As the implanted $Ge^+$ ion dose increases from 1.50 to 2.25 to 3.00 × $10^{16}$ $cm^{-2}$, the peak Ge concentration increases from 36% to 55% to 63%. Additionally, samples with higher $Ge^+$ ion doses have thicker SiGe layers. By asserting that the SiGe layer terminates when the Ge concentration drops below 10%, the thicknesses of samples A, B, and C are estimated at 4.6, 7.0, and 9.1 nm, respectively. We note that additional O and C signals were observed at the surface and were assumed to be the result of surface native oxide and contamination, respectively. The region of surface native oxide is not shown in Figure 1 (see supporting information [SI] for additional AES information and plots showing O and C signals).

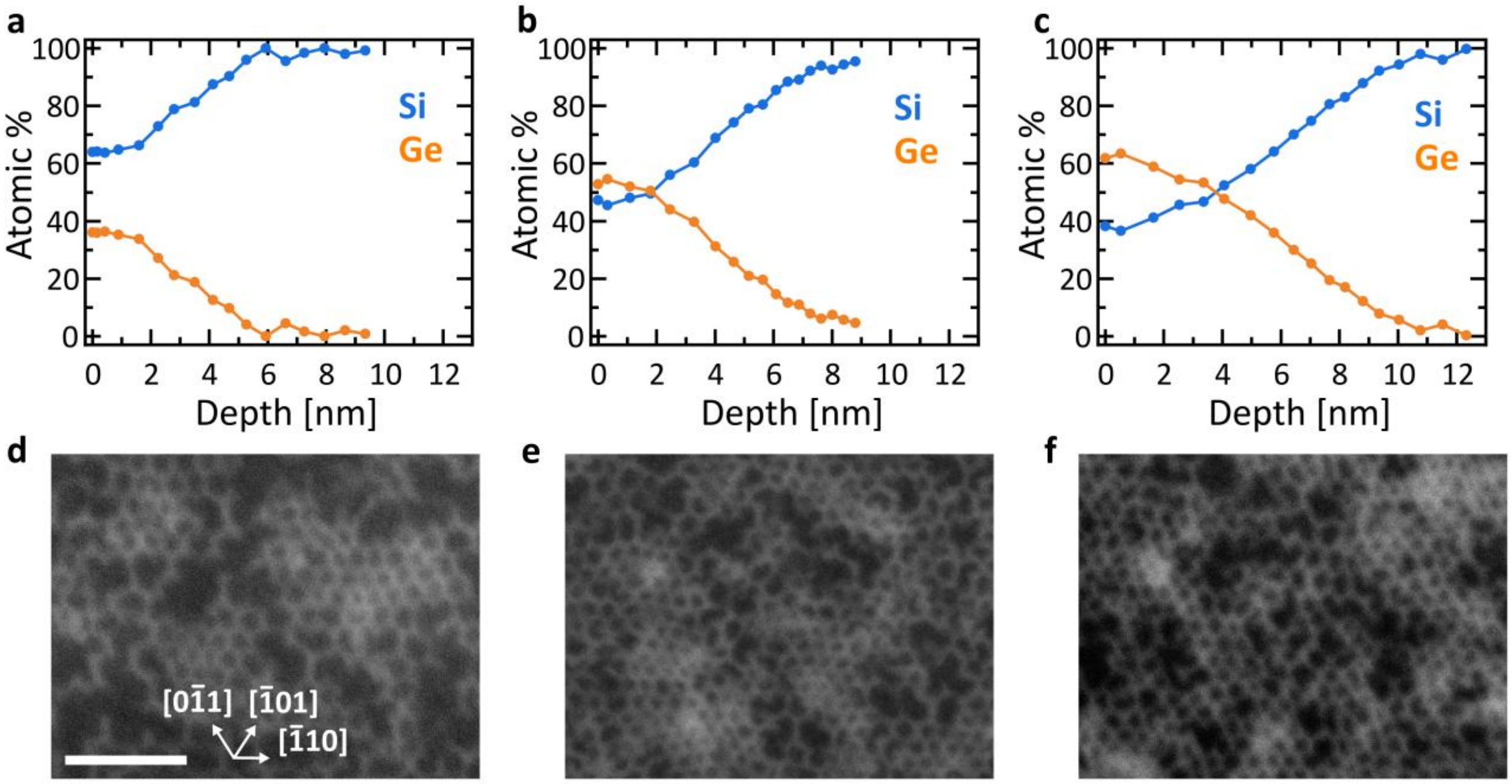

**Figure 1. a**–**c**) AES concentration depth profiles of Si and Ge for samples A–C, respectively. Implanted Ge concentrations are (**a**) 1.5 x $10^{16}$ cm$^{-2}$, (**b**) 2.25 x $10^{16}$ cm$^{-2}$, and (**c**) 3.0 x $10^{16}$ cm$^{-2}$. **d**–**f**) Top-view SEM micrographs of samples A–C, revealing a honeycomb web of contrast across the sample surface. The scale bar in (**d**) corresponds to 100 nm. Scale bar and direction vectors in (**d**) also apply to (**e**–**f**).

Plan-view SEM images of samples A–C are shown in **Figure 1d–f**, revealing an irregular hexagonal pattern of light lines enclosing darker regions. Atomic force microscopy (AFM) topographs show no indication of height variations resembling this pattern, implying the SEM contrast is not due to topography (see SI). Qualitatively, the spacing of the features (referred to as the hexagonal pattern) decreases and becomes more uniform with increasing Ge implantation dose. The pattern spacing and uniformity are analyzed quantitatively below. We note that this hexagonal pattern is consistent with dislocation models for SiGe on Si(111), which can exhibit hexagonal dislocation networks with dislocation lines running along $\langle 1\bar{1}0 \rangle$ directions.[14,15,21,22] To further explore the origin of these SEM features, plan-view TEM was performed on an additional sample with $Ge^{+}$ ion implantation dose of $3.00 \times 10^{16}$ cm$^{-2}$ (sample D).

**Figure 2a** presents an SEM image of sample D, illustrating a similar hexagonal pattern of light and dark regions to those seen in Figure 1, with an estimated spacing of 15 ± 1 nm. A high-angle annular dark field scanning transmission electron microscopy (HAADF-STEM) image of this sample is shown in **Figure 2b**, depicting a pattern of dark contrast spots spaced 15.1 ± 0.5 nm apart, aligned along $\langle \bar{1}10 \rangle$. Given that the sizing and orientation of the pattern seen in HAADF-STEM matches the features seen in SEM (c.f. Figure 2a–b), we conclude that they correspond to the same nanostructuring. A STEM energy dispersive x-ray spectroscopy (STEM-EDS) map of the Ge concentration for sample D is shown in **Figure 2c**, corresponding to the same region as the STEM image in Figure 2b. The EDS map indicates that the bright regions in the HAADF and SEM micrographs correspond to higher local concentrations of Ge. The symmetry and linearity of this pattern suggests that the periodic composition variation is related to a network of dislocations at the SiGe/Si(111) interface. This is consistent with our recent work suggesting that defects at the SiGe/Si interface facilitate the enhanced diffusion of Ge during oxidation.[20]

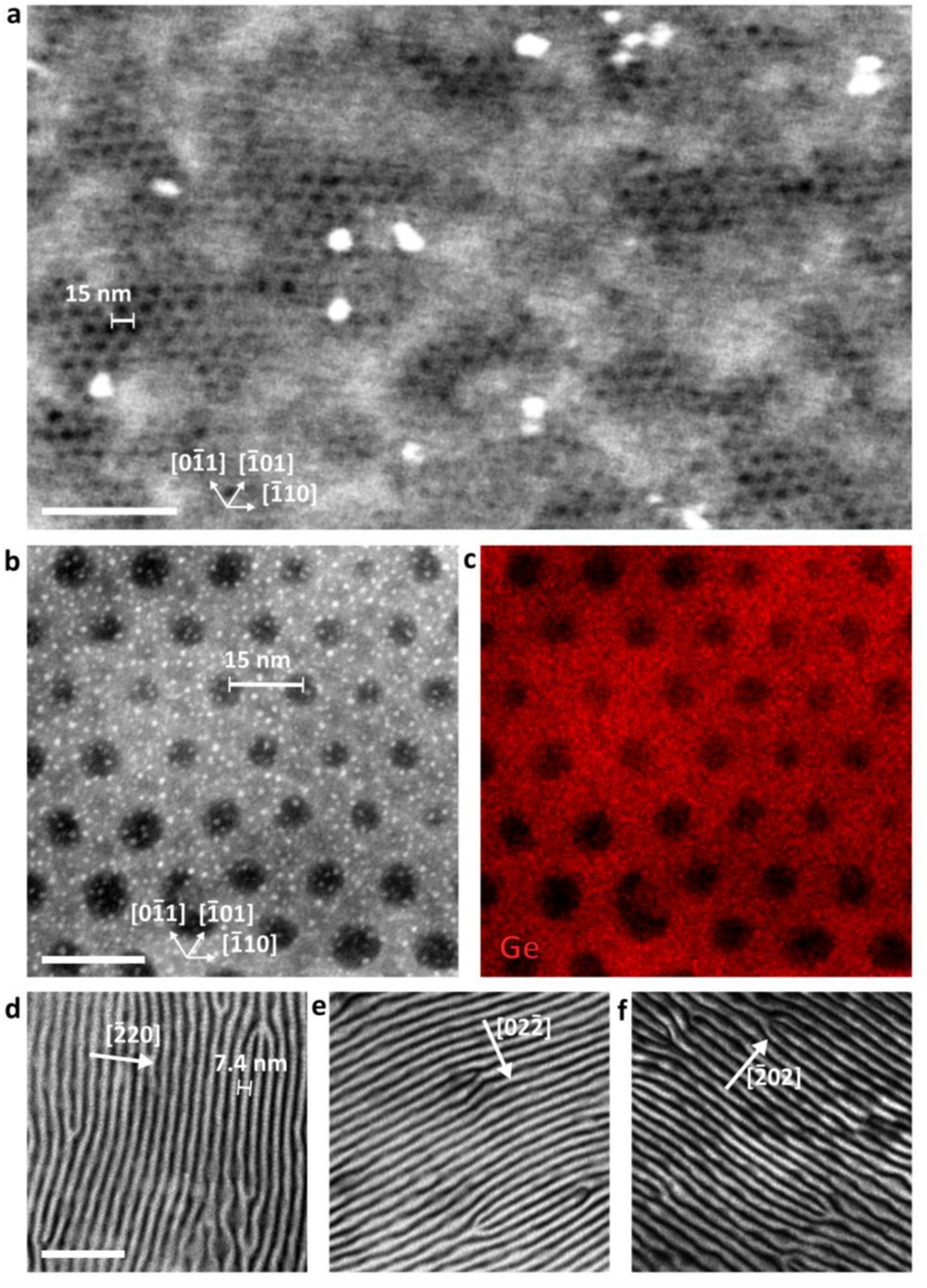

**Figure 2. a**) Plan-view SEM of sample D with scale bar corresponding to 100 nm. **b**) Plan-view HAADF-STEM image of sample D with scale bar corresponding to 20 nm. **c**) EDS composition map of Ge concentration taken on the same area as in (**b**). **d**–**f**) Dark-field TEM images of sample D. In each image, the sample is tilted off axis toward one of the three in-plane $\langle\bar{2}20\rangle$ directions, as indicated on each figure. Scale bar in (**d**) corresponds to 50 nm and applies to **e**–**f**).

**Figures 2d–f** display dark field off-axis TEM micrographs for sample D, tilted toward the three symmetric in-plane $\langle\bar{2}20\rangle$ directions as indicated in each panel. The dominant contrasts observed are Moiré fringes, implying strain relaxation in the SiGe layer. The fringes are spaced 7.4 ± 0.4

nm apart, which corresponds to a 2.6 ± 0.2% difference between the in-plane lattice spacings of the SiGe and Si layers. This difference is sufficient to fully relax a SiGe film containing 62 ± 3% Ge. We note that the Auger depth profile for this sample (shown in the SI) exhibits a peak Ge concentration of 80%, suggesting the film is at least 77% relaxed (62% / 80%) based on the spacing of the Moiré fringes. A proposed dislocation network responsible for the observed Moiré pattern and hexagonal composition nanopatterning is discussed below.

Strain relaxation and dislocation networks in SiGe/Si(111) heterostructures have been explored previously.[14,15,21–25] In some of these works, the addition of an Sb surfactant was used to prevent Ge islanding. For SiGe(111) grown with the Sb surfactant, rather than nucleating at island edges—as is the case for growth without the surfactant—Shockley partial dislocations nucleate at the surface of the film, gliding to the interface with the half loop bound by threading dislocations.[14,15] A second Shockley partial dislocation is energetically favored to nucleate at the intersection of the surface and the first threading dislocation. The second partial dislocation then glides to the interface along the same plane as the first dislocation, annihilating the first threading dislocation. The result is a SiGe film with a network of dislocations at the interface that is theoretically free of threading dislocations. The interfacial dislocations have lines running in the threefold symmetric $\langle \bar{1}, 1, 0 \rangle$ directions and are found in pairs of partial dislocations—one 30° and one 90°—with $\langle \bar{2}, 1, 1 \rangle$ Burgers vectors. Each pair of partial dislocations is equivalent to one full 60° dislocation, with $\langle \bar{1}, 1, 0 \rangle$ Burgers vectors. The pairs of partial dislocations and their equivalent full dislocations have the following Burgers vectors:

$$\frac{1}{6}\langle \bar{1}, \bar{1}, 2 \rangle + \frac{1}{6}\langle \bar{2}, 1, 1 \rangle = \frac{1}{2}\langle \bar{1}, 0, 1 \rangle$$

$$\frac{1}{6}\langle 1, 1, \bar{2} \rangle + \frac{1}{6}\langle \bar{1}, 2, \bar{1} \rangle = \frac{1}{2}\langle 0, 1, \bar{1} \rangle$$

$$\frac{1}{6}\langle 2, \bar{1}, \bar{1} \rangle + \frac{1}{6}\langle 1, \bar{2}, 1 \rangle = \frac{1}{2}\langle 1, \bar{1}, 0 \rangle$$

For a detailed explanation of the formation of these dislocations, the reader is referred to LeGoues et al.[15] LeGoues's study presents bright-field TEM images with similar Moiré fringes to those of Figure 2 above. Furthermore, the direction of the dislocations formed in this study matches the orientation of the patterns seen in SEM and HAADF-STEM of our samples, suggesting that the observed micrograph patterns are the result of the same dislocations. Next, we quantitively compare the observed patterns with the proposed dislocation model.

The spacing of strain-relieving dislocations is directly related to the in-plane strain relaxation between the film and the substrate—greater relaxation requires more closely packed dislocations. Assuming strain is relaxed via the pairs of partial dislocations discussed above, we can calculate the expected dislocation pair spacing corresponding to the 2.6 ± 0.2% mismatch deduced from the Moiré pattern (Figure 2d-f). The total lattice distortion in any direction on the (111) plane is a sum of the distortion caused by all three pairs of partial dislocations. Using the equivalent full dislocations simplifies the calculation, as there are only three Burgers vectors to analyze as opposed to six. For relaxation in the $[1, 1, \bar{2}]$ direction, the distortion from each type of full dislocation is found by taking the dot product of the Burgers vector with the $\langle 1, 1, \bar{2} \rangle$ unit vector, then multiplying by the lattice constant of Si. Accordingly, the three full dislocations—one of each type—cause a net distortion of 0.665 nm. Dividing this by the 2.6 ± 0.2% in-plane lattice mismatch, a predicted full-dislocation spacing of 25.6 ± 0.9 nm is found. This results in a predicted partial dislocation spacing of 12.9 ± 0.5 nm required to achieve the strain relaxation deduced from the Moiré fringes.

**Figure 3a** presents a schematic of the proposed dislocation array from LeGoues,[15] and **Figure 3b** superimposes this pattern on top of the HAADF-STEM image, matching the dislocation spacing with the observed pattern. The dislocation model matches the HAADF-STEM image for a partial dislocation spacing of 13.1 ± 0.4 nm. This agrees with the predicted partial dislocation spacing of 12.9 ± 0.5 nm, calculated above from the Moiré pattern. The quantitative agreement between the two provides compelling evidence that the observed patterns seen in HAADF and SEM correspond to the proposed network of interface dislocations. Comparing Figure 3b with the EDS map of Figure 2c, we see the dislocation lines correspond to regions of locally high Ge concentration. EDS maps of Si over the same area (shown in SI) confirm these high-Ge lines encompass regions that are rich in Si. This dislocation-driven composition nanostructuring agrees with our recent report of defect-mediated diffusion during SiGe oxidation driving interface composition fluctuations in the SiGe/Si(111) system. It was proposed that diffusion of Si and Ge is enhanced in the vicinity of the dislocations, enabling the SiGe condensation layer to penetrate deeper into the substrate near dislocations during oxidative SPE.[20]

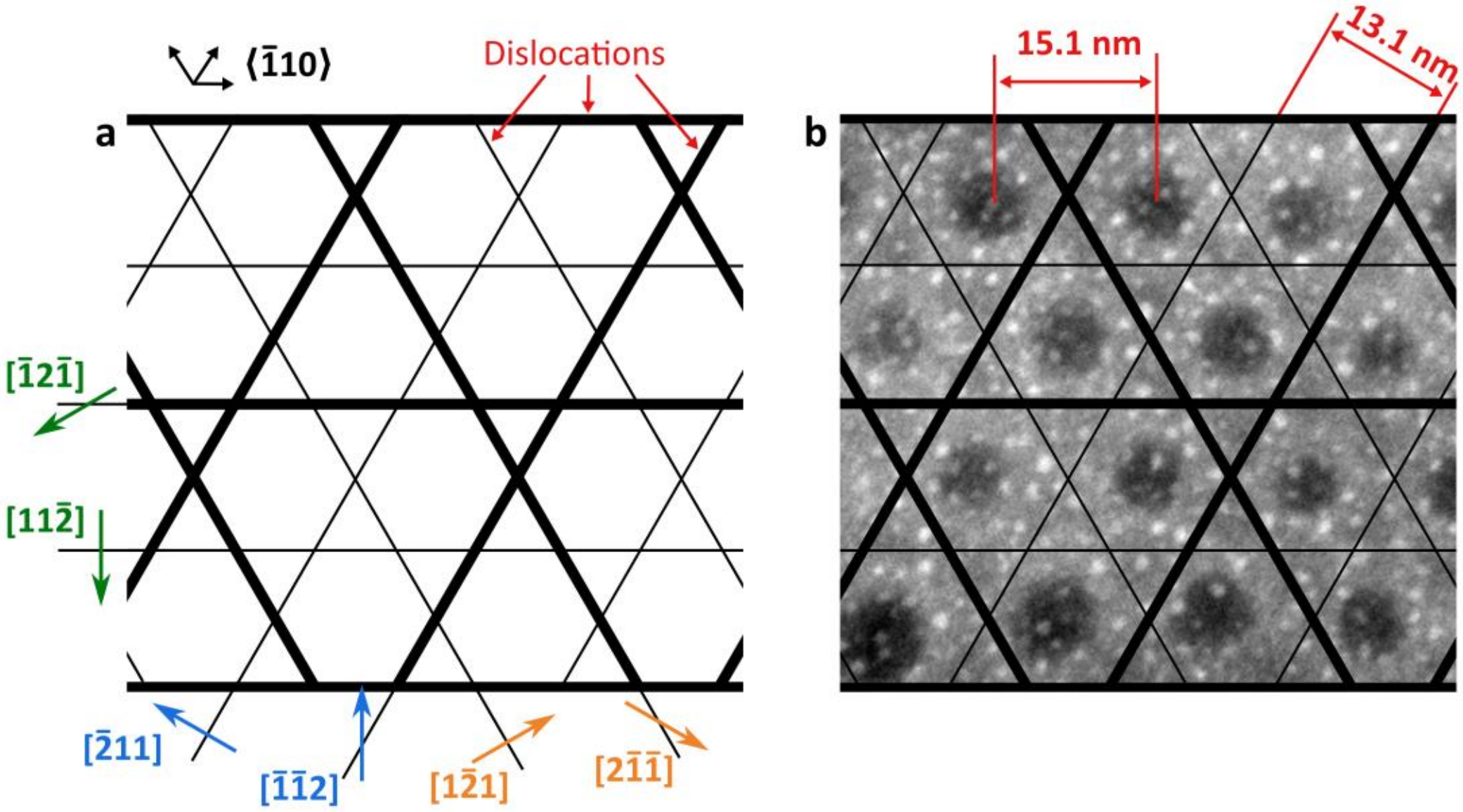

**Figure 3. a**) Schematic of interface dislocations for SiGe on Si(111) consistent with LeGoues.[15] The bold lines represent 30° partials and thin lines represent 90° partials. The triangular regions formed by the intersection of dislocations correspond to stacking faults bordered by Shockley partial dislocations. **b**) The proposed dislocation pattern overlaying the HAADF-STEM image from Figure 2.

The above discussion illustrates that the hexagonal pattern observed in SEM/TEM is the result of composition nanostructuring caused by a pattern of strain-relieving partial dislocations at the SiGe/Si interface. Therefore, the observed pattern in SEM provides the opportunity to estimate the local in-plane lattice parameter of the SiGe film, offering a simple method to map the local strain state (in-plane lattice constant) across the sample using a fast and non-destructive characterization method. Such strain information is especially useful for subsequent heteroepitaxy, where knowledge of the in-plane lattice spacing is essential. To quantify film relaxation, we developed an image-based pattern recognition algorithm to characterize the hexagonal pattern seen in SEM and estimate resulting film properties.

**Figure 4** illustrates the pattern recognition algorithm output for an SEM image of sample B. Figure 4a shows an SEM image that is fed into the algorithm, and Figure 4b presents the algorithm's output. The green shapes in Figure 4b enclose the detected dark regions that correspond to Si-rich pockets, and the blue lines indicate the nearest neighbour distances between these pockets. The dislocation spacing is $\sqrt{3}/2$ times this distance (see SI). Figures 4c–d depict enlarged regions from Figures 4a–b, respectively. We note that there are regions in the images where the code is unable to detect a pattern, the implications of which are discussed below.

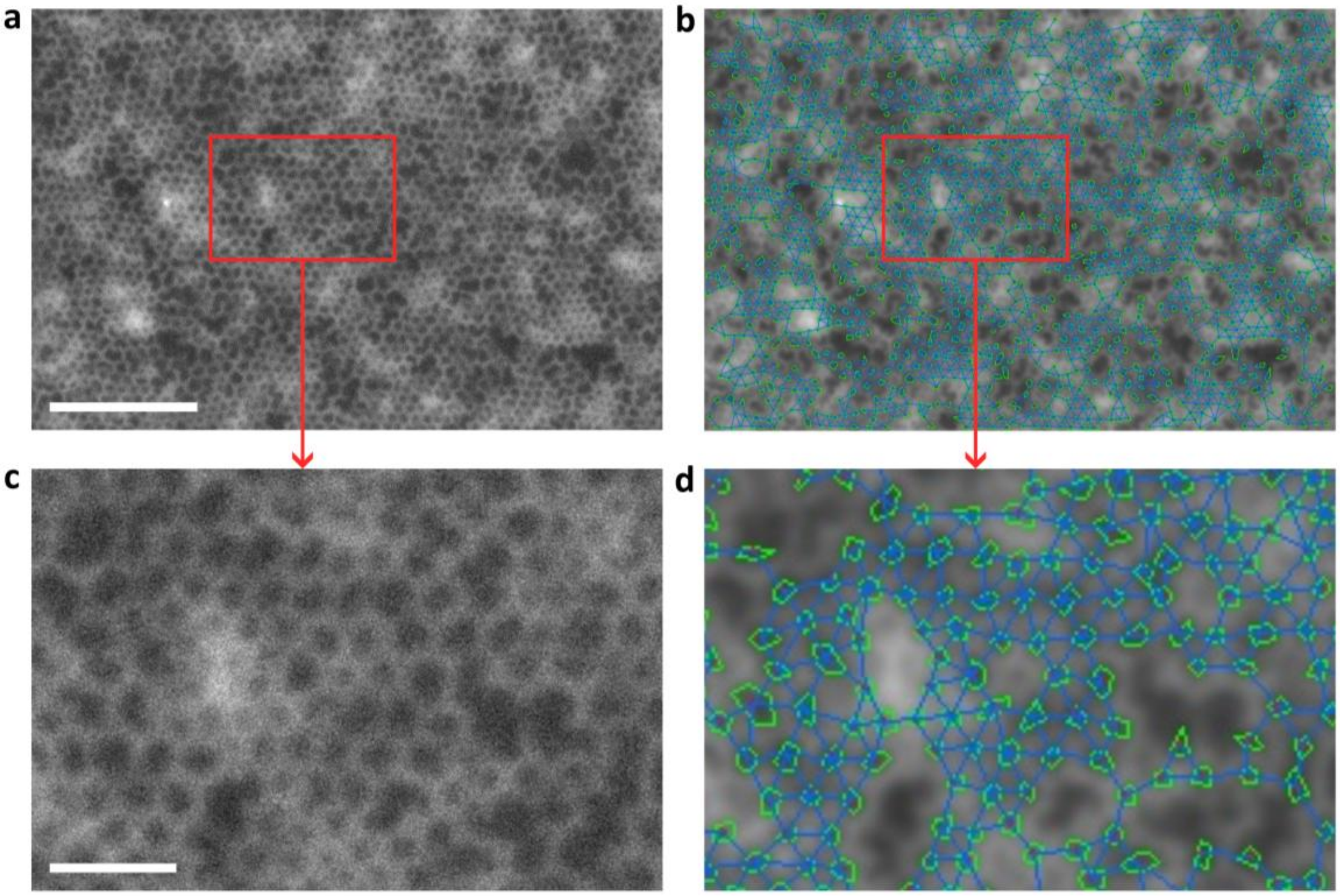

**Figure 4. a**) SEM image from sample B with 200 nm scale bar. **b**) The pattern detected by the algorithm overlaid on the SEM image from (**a**). Green lines enclose dark regions (of low Ge concentration), and blue lines indicate the distance between these regions (i.e. pattern length). **c**–**d**) Enlarged areas of figures (**a**) and (**b**), respectively, corresponding to the red box in each image. The scales bar in (**c**) is 50 nm and also applies to (**d**).

**Figures 5a–d** show histograms of partial dislocation spacings deduced by the algorithm for samples A–D, respectively. Every count on the histograms corresponds to a measured partial dislocation spacing detected from the SEM image by the algorithm. Gaussian fits to the histograms are also displayed in the figures. Comparing samples A–C (implanted with 1.50, 2.25 and 3.00 × $10^{16}$ $cm^{-2}$ $Ge^{+}$ at the same implantation energy, Figures 5a–c), it is seen that increasing the $Ge^{+}$ dose decreases both the dislocation spacing and the full width half max (FWHM) of the distribution. The smaller dislocation spacing is expected due to the larger SiGe–Si lattice mismatch for samples with higher Ge concentrations, resulting in higher dislocation formation and strain relaxation. Furthermore, the narrowing of the FWHM indicates higher uniformity of the dislocation network at larger $Ge^{+}$ doses, consistent with visual inspection of the SEM images in Figure 1.

`

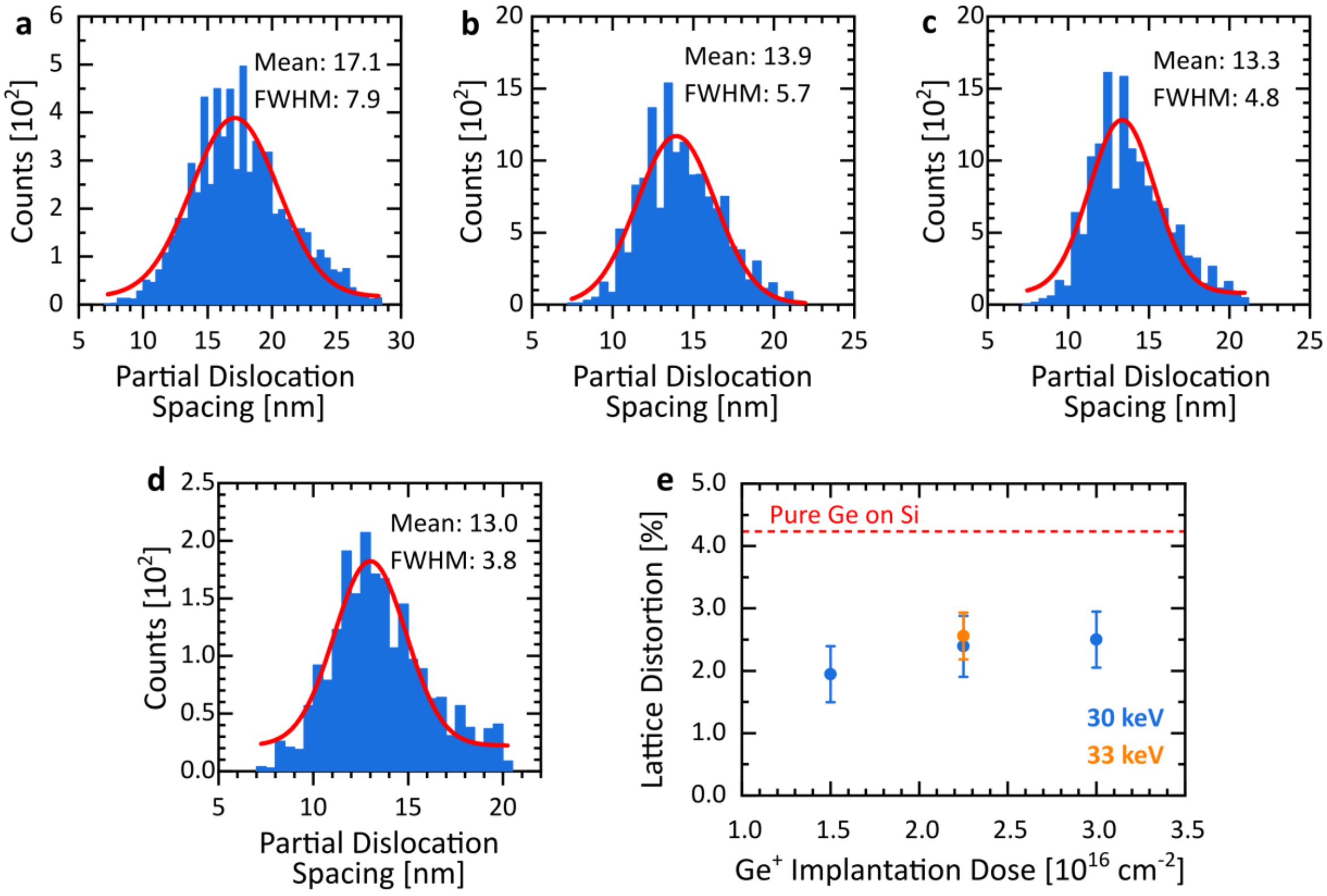

**Figure 5**. **a–d**) Histograms of the partial dislocation spacings found in samples A–D, respectively, using the algorithm. **e**) The in-plane lattice distortion deduced for each sample, compared to the relaxation of pure Ge on Si (shown as the red dotted line). The legend indicates the ion implantation energies of each data point.

**Figure 5e** presents the relative in-plane lattice distortion of the SiGe films relative to Si for samples A–D, as deduced from the mean dislocation spacings output from the algorithm. The lattice distortion for each film is calculated assuming the mean dislocation spacing is applicable to the entire surface. Using this assumption for sample D, we calculate a lattice distortion of 2.6 ± 0.4% which remarkably matches the lattice distortion found from the TEM Moiré Fringes. As mentioned above, the algorithm could not detect a pattern on the entire SEM image area. While it is possible that these undetected regions exhibited different strain relaxation from the detected regions, the agreement of strain relaxation between the TEM and SEM data for sample D provides compelling evidence that there is no significant discrepancy. This aligns with observations that the hexagonal pattern is uniform and extends across the entire imaged region of the TEM specimen, despite the pattern only being visible in about 50% of SEM image for this sample (c.f. Figure 2a–b).

The plot in Figure 5e shows that the in-plane lattice distortion increases with increasing $Ge^+$ dose. For samples A–C the distortion increases from 1.9 ± 0.4%, to 2.4 ± 0.5%, to 2.5 ± 0.5%, respectively. By comparison, a relaxed pure Ge layer on Si would have 4.2% distortion, displayed as a dotted red line in Figure 5e. Therefore, the amount of lattice distortion occurring in each sample is equivalent to fully relaxed SiGe films with Ge concentrations of 45 ± 9%, 57 ± 12%, and 60 ± 12% for samples A–C, respectively. We note that the peak Ge concentrations in the AES profiles were 36%, 55%, and 63%, suggesting the SiGe layers may have between 80 and 100% strain relaxation. This insight into local lattice distortion of SiGe layers can be invaluable for subsequent heteroepitaxial growth of III-Vs.

## 3. Conclusion

We have explored the highly efficient strain relaxation process in SiGe(111) metamorphic layers formed by oxidative SPE. SiGe layers about 10 nm thick and containing up to 80% Ge exhibit a hexagonal nanopatterning of the Ge composition. The compositional nanopatterning across the sample is caused by defect-enhanced diffusion of Si and Ge during the SPE process associated with a hexagonal network of dislocations at the SiGe/Si interface. A non-destructive method of characterizing the local strain state of the SiGe layer at the nanoscale is demonstrated by analyzing the SEM-observed pattern, which is validated by TEM data. The efficient strain relaxation of these layers, as well as the ability to characterize strain and strain uniformity by SEM, makes oxidative SPE a promising method for subsequent GaAs heteroepitaxy, presenting a novel approach for monolithic III-V integration on Si.

## 4. Methods

Four SiGe/Si(111) samples were prepared in this study. Si(111) wafers were first implanted with $Ge^+$ ions at 30–33 keV and various doping levels. Samples A, B, and C were implanted with $Ge^+$ ion doses of 1.5, 2.25, and 3.0 × $10^{16}$ $cm^{-2}$ at 30 keV. A fourth sample, D, was implanted with a dose of 3.0 × $10^{16}$ $cm^{-2}$ at 33 keV. After implantation, the samples were wet-oxidized in a tube furnace, where $O_2$ gas was bubbled through DI water maintained at 95 °C. Samples A–C were oxidized at 900 °C for 30 min, and sample D was oxidized at 850 °C for 60 minutes. The oxidation process selectively oxidizes Si atoms, producing a Ge-rich condensation layer at the interface between the (above) oxide and the (below) Si(111) substrate. The oxide layer was subsequently removed by submerging the samples in a buffered oxide etchant solution (10:1 NH4F:HF) for 5 minutes prior to sample characterization to remove the oxide layer.

SiGe surfaces were characterized by SEM using a Magellan 400 XHR microscope with a through-the-lens detector in immersion mode (to increase Z contrast). Concentration depth profiles of the samples were collected by AES on a JAMP-9500F Auger microscope, using an argon ion (ArIon) gun at 2 keV to etch into the surface of the samples. The electron beam was defocused to obtain an acquisition area of around 100 nm. To calibrate the etch rate of the ArIon gun, the depth profile of a similar SiGe/Si(111) sample was measured using both AES (with the same ArIon gun conditions) and EDS mapping in cross-sectional STEM. The depth from the EDS data was compared to the sputter time of the AES data to extract a four-order polynomial equation for the ArIon gun sputter rate as a function of Ge concentration. This equation was used to convert sputter time of the AES profiles to depth. For more detailed information on this calibration process, see the SI.

The microstructure of the SiGe layer and SiGe/Si(111) interface of sample D was characterized by plan-view high-resolution TEM and STEM-EDS. An FEI Titan 80-300 LB TEM equipped with a monochromator, hexapole aberration corrector, and double-tilt holder was used to image the sample. The sample was prepared using a Helios 5 UC focused ion beam (FIB) microscope using a $Ga^+$ ion source. Four sets of 10-nm-thick carbon horizontal tracking marks were deposited using electron beam deposition, each separated by a 20 nm thick deposited platinum layer, used as a reference for distance from the original specimen surface during thinning. Prior to lift-out of the surface-plane orientation, a protective 2 μm carbon layer was deposited over the entire surface. The lamella was attached to a copper FIB grid and re-mounted rotated 90°for milling to be conducted parallel to the original surface. A protective 1.5 μm tungsten layer was deposited over the new exposed capping surface, and the lamella was thinned to a thickness of approximately 70–80 nm at the location of the interface in the SiGe/Si film.

A pattern-recognition algorithm was developed to analyze the size of repeated features found in the SEM images. The algorithm detects the contours of closed polygons in a size range based on an initial estimate that is input by the user. Each detected polygon is then sorted using a k-nearest neighbors search algorithm to determine the Euclidian distance to the six nearest neighboring polygons. Distances are omitted from consideration if they are greater than 1.5 times the user-input size estimate. The average length and standard deviation of the remaining distances are calculated, and the distance is ignored if both the standard deviation is greater than 2.4 nm, and the distance is greater than average. This removes outlier distances that should not be connected. After sorting through all distances, the average of all the distances in the image is calculated and compared with the initial guess for pattern length. If the difference between the two values is greater than 0.01 nm, the algorithm recurses using the calculated distance as the updated pattern length guess. The algorithm continues recursing until convergence. See the SI for more details on the pattern recognition algorithm.

**Acknowledgements**

The authors are grateful to Doris Stevanovic and Nebile Isik Goktas for technical support, Travis Casagrande for AES support, Joseph Deering for FIB expertise, and Alexandre Pofelski for TEM assistance. The authors acknowledge support from McMaster University's Canadian Centre for Electron Microscopy and Centre for Emerging Device Technologies. The authors are grateful for financial support from the Natural Sciences and Engineering Research Council of Canada under projects RGPIN-2020-05721 and ALLRP-588298-23.

**Table of Contents**

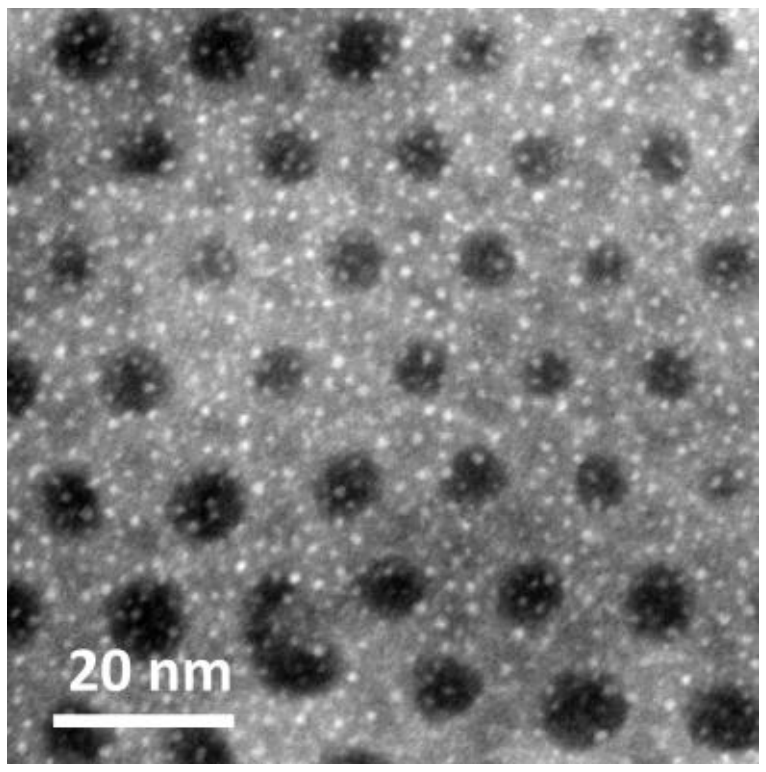

Thin SiGe buffer layers are grown by oxidative solid-phase epitaxy, resulting in a nanopatterning of varying Ge concentration across the surface of the sample. The nanopatterning is caused by a hexagonal network of dislocations on the SiGe/Si(111) interface, and is visible by scanning electron microscopy, providing a non-destructive method for obtaining local strain state information of the SiGe layer.

# Supporting Information

**Spontaneous Nanopatterning and Strain Relaxation in SiGe Layers Grown by Oxidative Solid Phase Epitaxy**

*Sophie E. Bierer, Trevor R. Smith, Arezoo Mafi, Sunzhuoran Wang, Chengqian Liao, Vatsalkumar Patel, Andrew P. Knights, Ryan B. Lewis**

The following supporting information provides additional auger electron spectroscopy (AES) depth profiles of all samples discussed in the main text—samples A, B and C implanted with $Ge^+$ doses of 1.50 × $10^{16}$, 2.25 × $10^{16}$ and 3.00 × $10^{16}$ $cm^{-2}$ at 30 keV, respectively, and sample D, implanted with 3.00 × $10^{16}$ at 33 keV. Also provided are Atomic force microscopy (AFM) scans of samples A–C, and scanning transmission electron microscopy energy dispersive x-ray spectroscopy (STEM-EDS) maps of sample D. Additionally, this text provides details on the methods use to derive the depth of the AES profiles from the sputter time, and further explains the pattern-recognition algorithm detailed in the main text.

**AES Depth Profiles**

**Figures S1a–d** show AES profiles of samples A–D for Si, Ge, O and C concentrations, plotted as a function of sputter time. The presence of O and C at the surface (zero sputter time) suggests these signals are the result of a surface oxide and surface contamination, respectively. The AES depth profiles in the main text have been cropped to remove all data points where the oxygen concentration is above 33%—approximately half the O concentration of pure $SiO_2$. The plots in the main text have also been normalized such that the Si and Ge signals sum to 100%. The depth profiles in Figure S1 show that samples A–D have maximum Ge concentrations of 32%, 46%, 51%, and 80%, respectively.

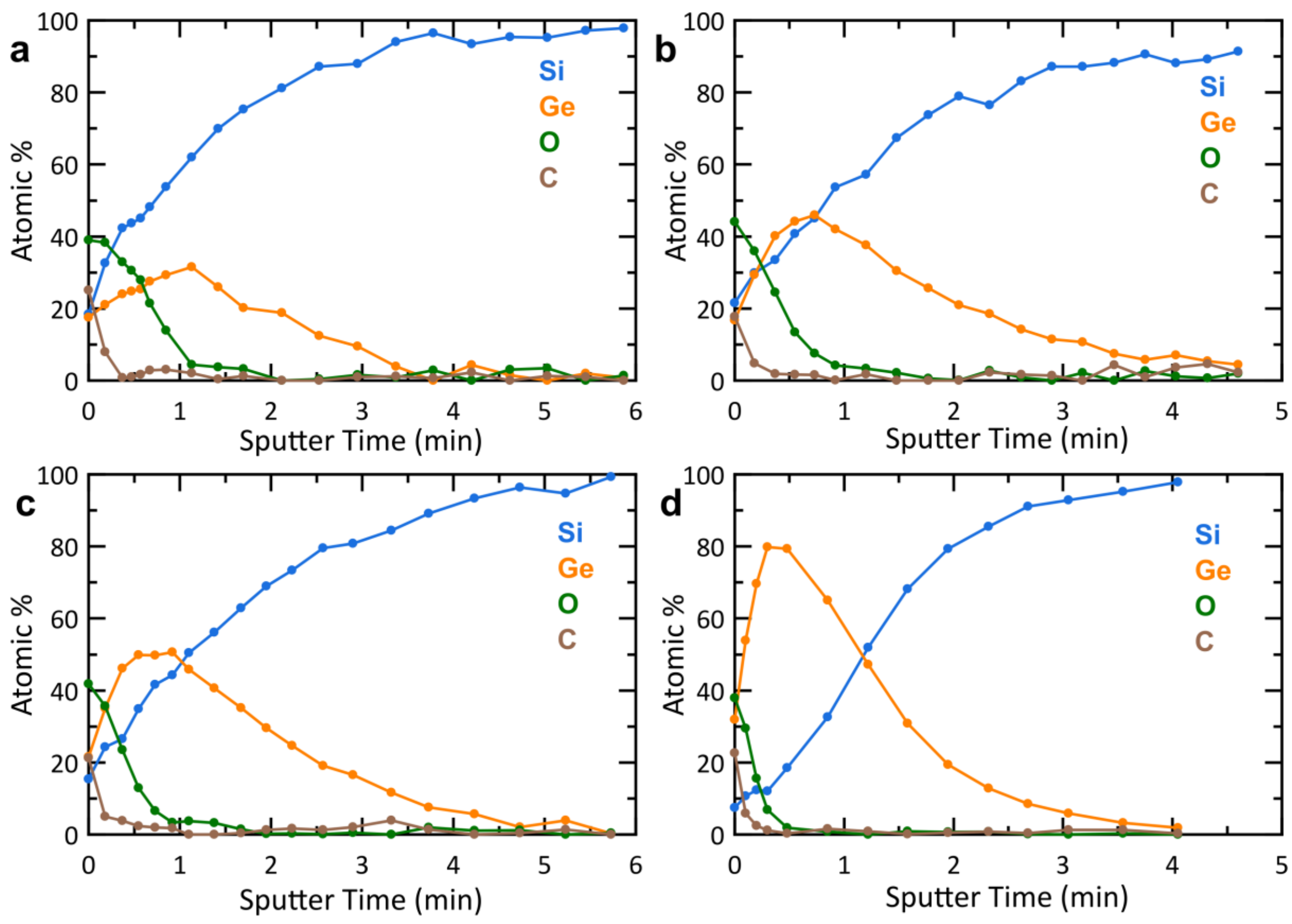

**Figure S1. a–d**) AES profiles plotted as function of sputter time for samples A–D, respectively

## AES Sputter Rate Calibration

To translate the AES sputter time into sputter depth, the sputter rate of the Argon Ion (ArIon) gun used for etching during the AES depth profiling was calibrated. For this, it was assumed that the most impactful variable factor impacting the sputtering rate is the Ge concentration at the sputtered surface of the SiGe film. To find a Ge-dependent sputter rate equation, a sample similar to samples A–D was characterized by combining two techniques: AES profiling (at the conditions used for samples A–D) and cross-sectional STEM-EDS. This additional sample was implanted with a $Ge^+$ dose of $3.00 \times 10^{16}$ $cm^{-2}$ at 35 keV. The STEM depth profile was acquired by integrating horizontally over a cross-sectional EDS scan. The two depth profiles used for this calibration are shown in **Figure S2**.

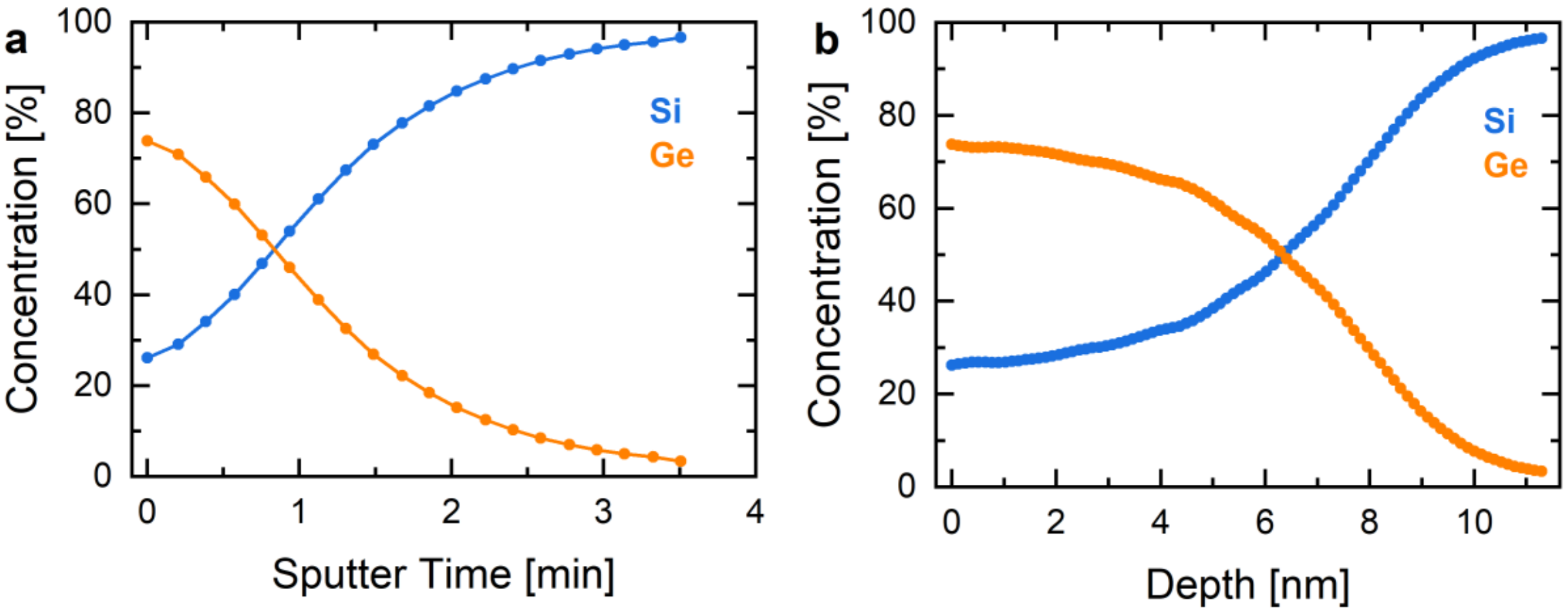

**Figure S2.** **a**) Normalized AES signals of Si and Ge as a function of sputtering time. **b**) STEM-EDS composition depth profile of the same sample as (**a**).

For each data point in the AES plot, a corresponding depth was interpolated from the EDS by matching the AES- and EDS-inferred concentrations. Doing so yielded three columns of data: sputter time ($t$), depth ($d$), and Ge concentration ($x$). The rate of change at each measured Ge concentration was approximated using a central-difference approach, as follows:

$$d'(x_i) = \frac{d_{i+1} - d_{i-1}}{t_{i+1} - t_{i-1}}$$

**Figure S3** shows the ArIon gun's sputter rate versus Ge concentration calculated using this method. The data was fitted to a fourth-order polynomial equation describing the sputter rate as a function of Ge concentration, displayed on the plot.

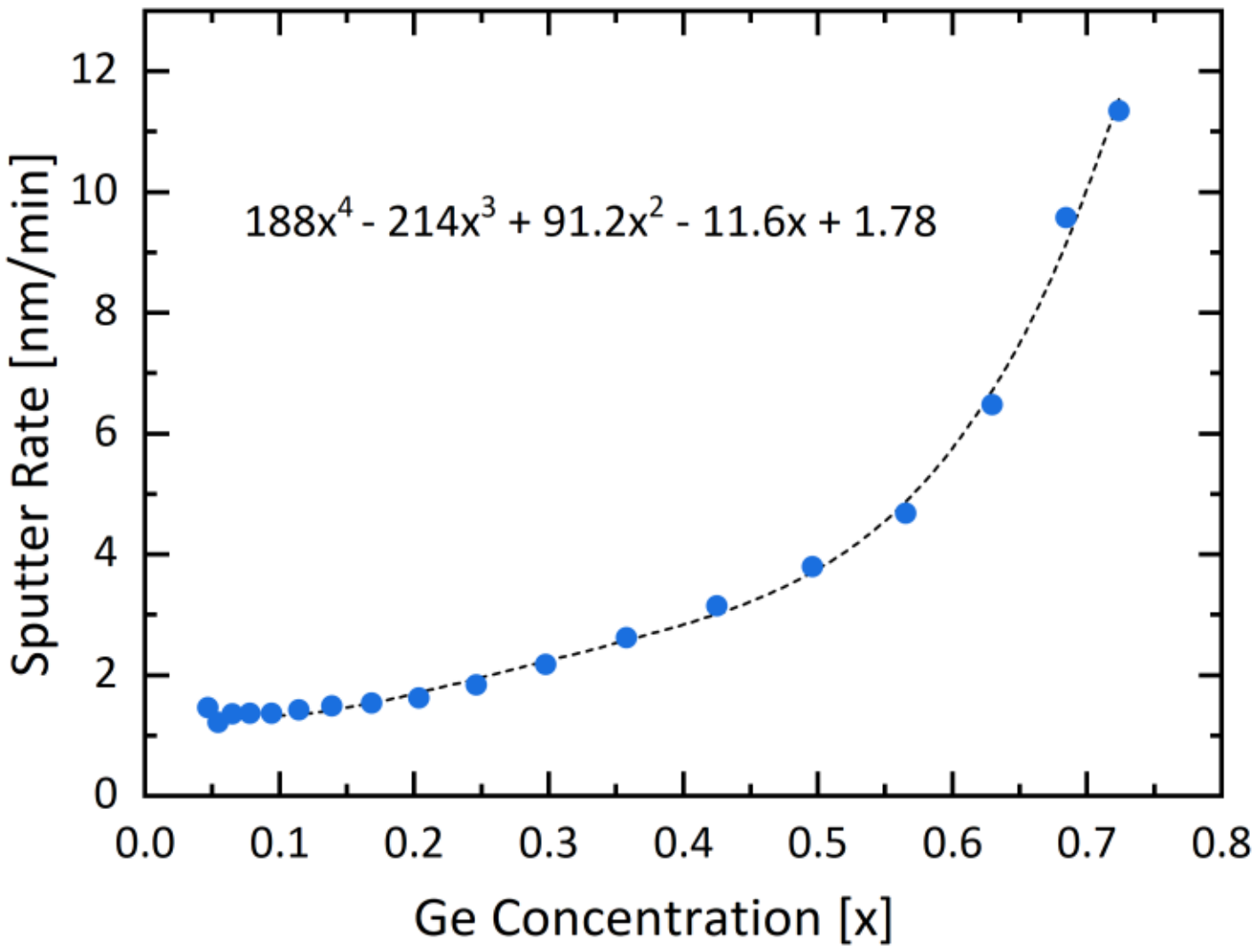

**Figure S3**. Calculated sputter rate versus Ge concentration and fourth-order polynomial fit.

To convert sputter time for the AES profiles in Figure S1 to depth, the AES data was first cropped to where O drops below 30% (as discussed above) then renormalized to only include Si and Ge. Once the data was cropped and renormalized, the fourth-order polynomial equation was used to calculate a sputter rate for each data point, then trapezoidal integration was used to approximate the depth, as shown below:

$$d_i = d_{i-1} + \left(\frac{d'(x_i) + d'(x_{i-1})}{2}\right) \times (t_i - t_{i-1})$$

**AFM**

Topographical AFM scans of samples A–C—collected using an Asylum Research MFP-3D AFM—are shown in **Figures S4a–c**. The raw AFM scans were levelled by mean-plane subtraction and rows aligned by a 3$^{rd}$ order polynomial. Scans of a 0.5μm by 0.5 μm area show no indication of the hexagonal pattern observed in SEM and TEM in the main text. Therefore, we conclude that the features seen in SEM are not topographical.

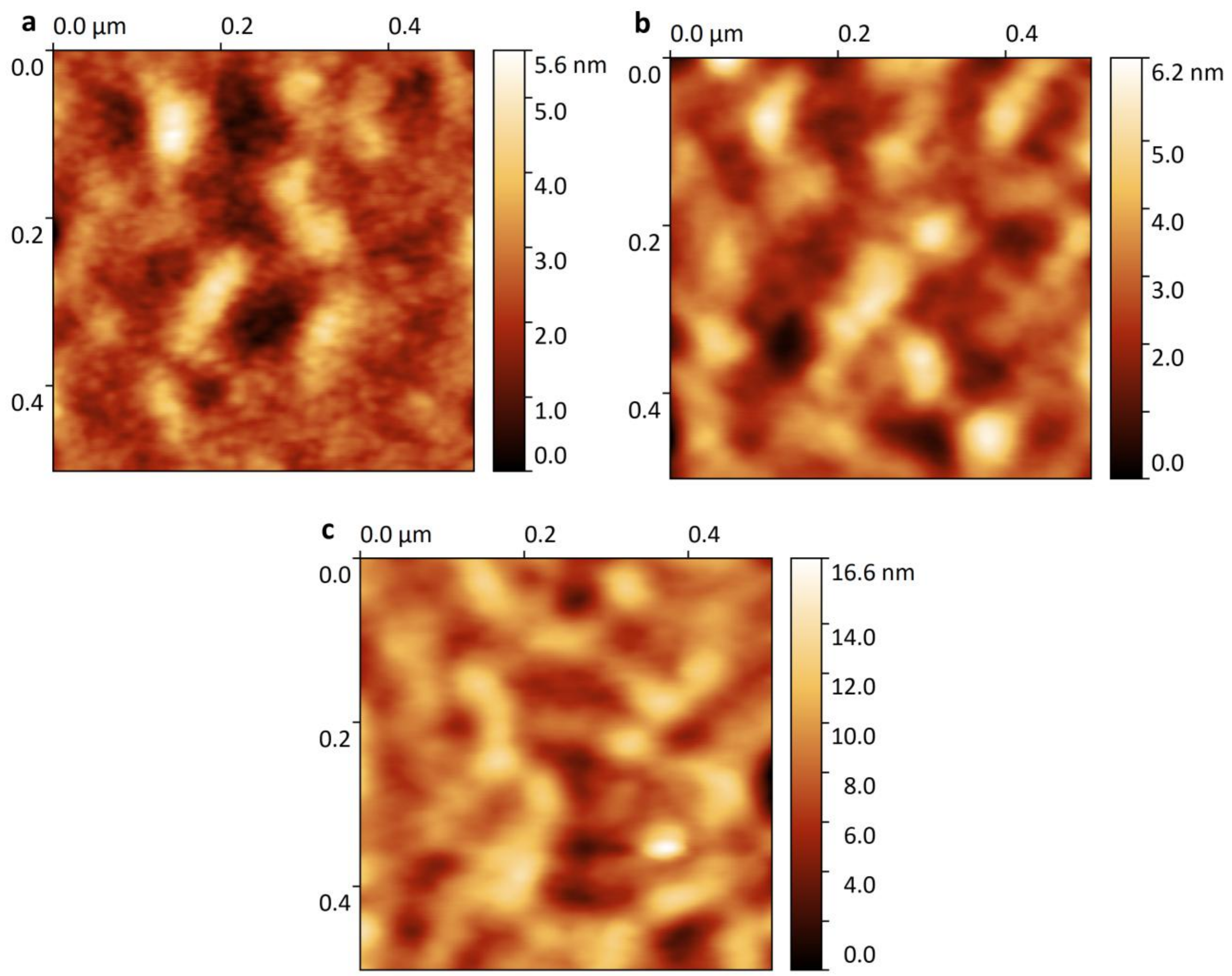

**Figure S4. a**–**c**) AFM topographs of samples A–C, respectively.

## EDS

**Figure S5a** depicts a TEM-EDS map of the Ge concentration across sample D, the same as that shown in Figure 2 of the main text. **Figure S5b** shows the TEM-EDS map of Si over the same area. These EDS maps show that regions of low Ge correspond to regions that are higher in Si. Though the contrast is not as pronounced in the Si map as it is in the Ge map, this is presumably due to the Si substrate saturating the signal.

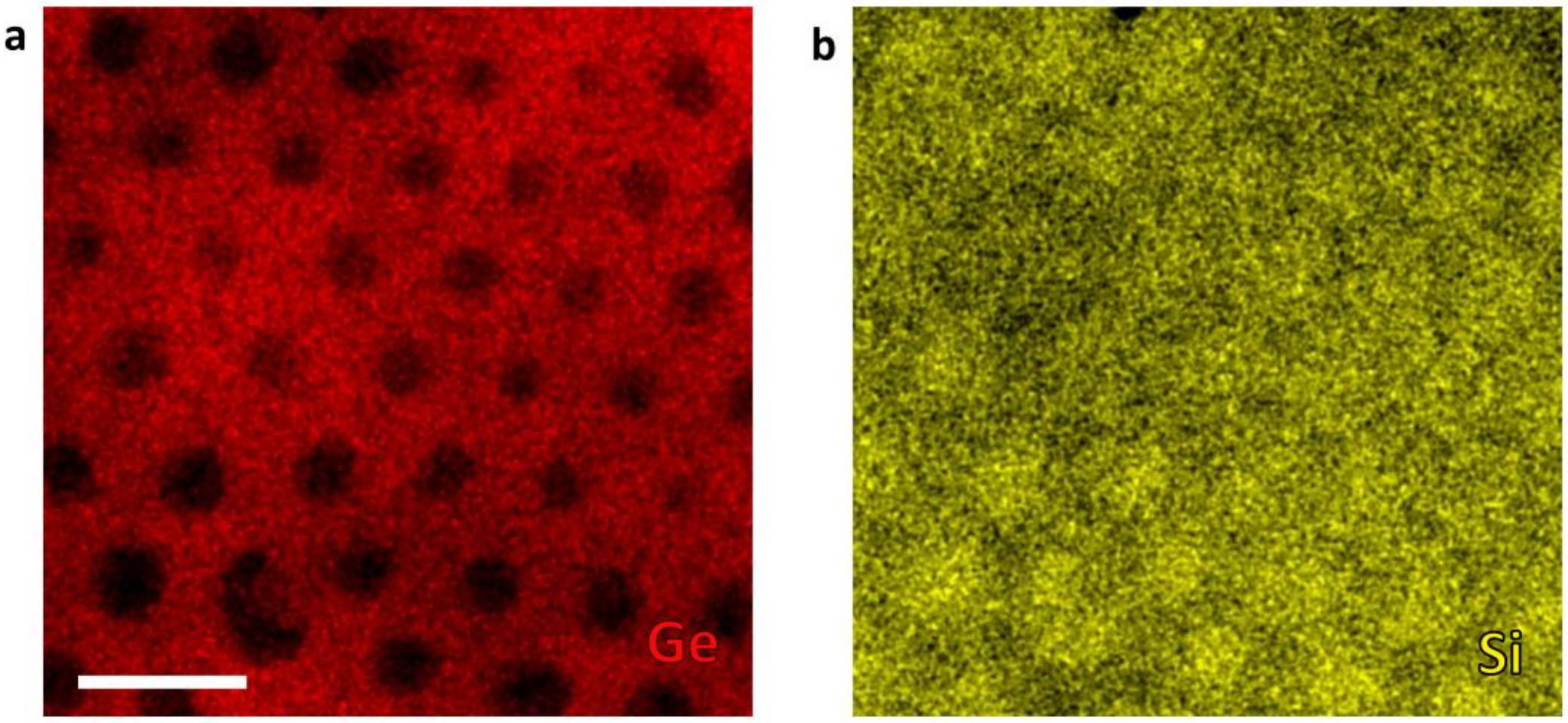

**Figure S5. a–b**) EDS maps from sample D showing Ge (**a**) and Si (**b**) EDS intensity. The scale bar corresponds to 20 nm

## Pattern-Recognition Algorithm

A pattern-recognition algorithm was developed to analyze the spacing of the hexagonally packed dark (Si-rich) regions in the SEM images. The algorithm was developed using Python's NumPy and OpenCV libraries. Pseudocode for the algorithm is shown in **Tables S1–S3**. **Figure S6** outlines how the algorithm detects these hexagonally packed, Si-rich regions in an SEM image, and how it approximates the dislocation spacing from the length of the hexagonal pattern.

First, an SEM image at 350,000x magnification is fed into the algorithm. Figure S6a depicts one such SEM image from sample D (this is a cropped version of the SEM shown in Figure 2a in the main text). The algorithm first detects contours in the image, shown in Figure S6b. Not all the detected contours correspond to the hexagonal pattern, so a filtering process analyzes each detected contour to remove those deemed too large or too small. For more details on this filtering process, see lines 48–66 of the pseudocode, shown in Table S2. The remaining contours are outlined in green in Figure S6c, which are enlarged in Figure S6d. These regions are referred to as hexagons due to their hexagonal packing, despite not necessarily being perfect six-sided polygons.

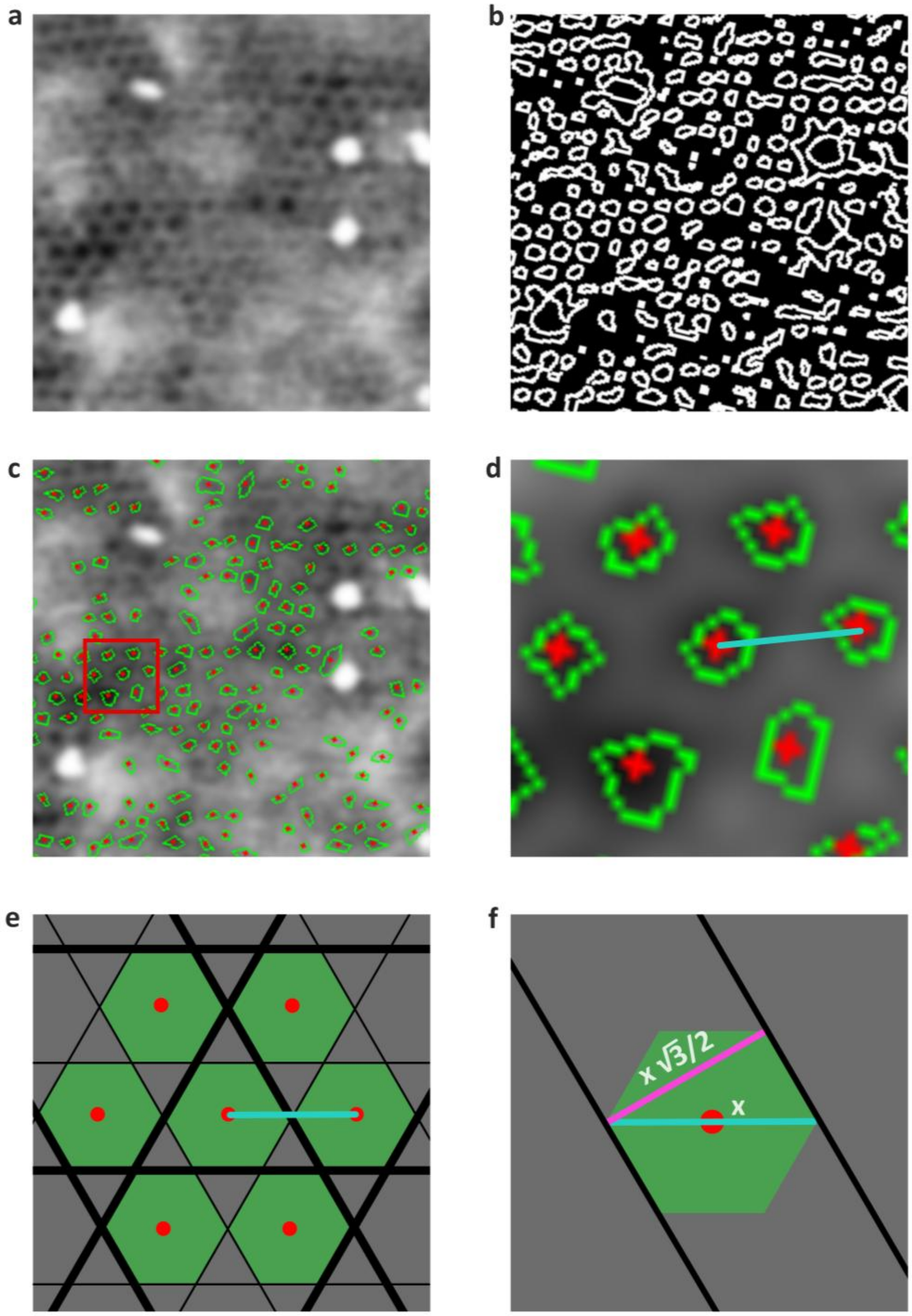

**Figure S6. a**) SEM of sample D. **b**) Contours detected in the SEM image by the algorithm. **c**) Original SEM image with the filtered contours outlined in green. Each red dot represents the center of these contours. **d**) enlarged version of the red box in (**c**). The blue line represents one pattern length/hexagon size measurement. **e**) The dislocation model, with thick and thin lines to show the alternating 30° and 90° partial dislocations. The green hexagons correspond to green contours in (**d**). **f**) A simplified dislocation model to show the relation between pattern length/hexagon size (blue line) and the dislocation spacing (pink line).

The distances between every detected hexagon are recorded, noting that one hexagon can only have up to six nearest neighbors. Additionally, not all hexagons have six detected nearest neighbors, as there are regions in the SEM image where the hexagonal pattern is not visible. Therefore, the algorithm undergoes another filtering process to determine whether each recorded distance corresponds to a neighboring hexagon—see lines 103–113 in the pseudocode (shown in Table S3) for details of this filtering process. In the discussion and pseudocode, these distances are interchangeably referred to as line length, hexagon size, or pattern length. Once every hexagon size is recorded, it is converted to a partial dislocation spacing by multiplying by a factor of $\sqrt{3}/2$, as depicted in Figure S6f. The final output of the algorithm is a list of every recorded partial dislocation spacing. These lists were generated for every sample, then plotted in histograms in the main text. The histograms were used to analyze uniformity and lattice distortion, as detailed in the main text.

**Table S1.** Pseudocode of the algorithm's main function

```
    ALGORITHM: SEM pattern recognition & analysis
    FUNCTION: main()
    Inputs
        img (matrix)... from grayscale SEM image, M x N array of intensity pixels
        est (float).... estimate for hexagon size
    Outputs
        avgDis (float).. average partial dislocation spacing
        disList (list).. every recorded dislocation spacing
    ----------------------------------------------------------------------------
1   avgHex (float)
2   diff (float) ← 100.0
3   (list) ← [] //List of hexagon sizes
4   disList (list) ← [] //list of dislocation spacings
5
6   WHILE diff < 0.01
7       //Step 1: Process the image
8       img_cropped (matrix) ← crop img to remove scalebar
9       img_smooth (matrix) ← apply gaussian blur to img_cropped
10      outline (matrix) ← contours of img_smooth outlined in b&w
11      contours (list) ← list of vertex coordinates for every contour in outline
12          //Use adaptiveThreshold() and Canny() from cv2 to get outline
13          //Use findContours() with RETR_LIST from cv2 to get contours
14
15      //STEP 2: Detect Si-Rich regions (aka hexagons)
16      centroids (list) ← DETECT_HEXAGONS(contours, est)
17
18      //STEP 3: Connect the hexagons to nearest neighbors
19      linePoints (list) ← FIND_NEIGHBORS(centroids, est)
20          //linePoints is a list of [(x1,y1),(x2,y2)] where (x,y) represents the\
21              //start and endpoint coordinatjies of each nearest neighbor line
22
23      //STEP 4: Find average hexagon size
24      FOR i IN RANGE 0 TO length(linePoints)
25          x1, y1 ← linePoints[i][0][0], linePoints[i][0][1]
26          x2, y2 ← linePoints[i][1][0], linePoints[i][1][1]
27          lineDist ← sqrt((x2 – x1)^2 + (y2 – y1)^2)
28          APPEND lineDist TO hexDist
29      avg_hex ← sum(hexList) / length(hexList)
30
31      //STEP 5: Check if convergence is reached
32      diff ← abs(hex – est)
33
34  //STEP 6: Convert every hex size to a dislocation spacing
35  disList (list) ← []
36  FOR hexSize IN hexList
37      disSpacing ← hexSize * sqrt(3) / 2
38      APPEND disSpacing TO disList
39  avgDis ← avgHex * sqrt(3) / 2
40  avgDis_nm  ← convert avgDis from pixels to nm
41
42  RETURN avgDis, disList
```

**Table S2.** Pseudocode of the algorithm's helper function, detect_hexagons()

```
   ALGORITHM: SEM pattern recognition & analysis
   HELPER FUNCTION: DETECT_HEXAGONS()
   Inputs
      contours (list)... list of vertex coordinates for every contour of SEM
      est (float)... estimate for hexagon size
   Outputs
      centroidsClean (list)... coordinates of centers of every valid contour
   -----------------------------------------------------------------------------
43 hexList (list) ← [] //vertex coordinates of each valid hexagon
44 centList (list) ← [] //coordinate of centre of every hexagon
45 hexagonsClean (list) ← []
46 centroidsClean (list) ← []
47
48 //Step 1: Determine area thresholds
49 estArea (float) ← (est / 2)^2 * Pi
50 minArea (int) ← int(estArea)
51 minArea (int) ← int(0.1 * estArea)
52
53 //Step 2: Determine what contours are "valid" hexagons
54 FOR contour IN contours
55    approx (list) ← simplified contour
56       // the contour is simplified using approxPolyDP with\
57          // 0.04 * arclength(contour, TRUE) as epsilon
58    maxSideLength ← max side length of approx
59    IF maxSideLength > est
60       CONTINUE
61    IF length(approx) >= 4 //Check the simple contour has 4 or more vertices
62       area (float) ← area of current simple contour
63       IF area > minArea AND area < maxArea
64          [cX, cY] ← center coordinate of approx
65          APPEND approx TO hexList
66          APPEND [cx, cy] TO centList
67
68 //Step 3: Detect overlapping hexagons
69 n (int) ← length(hexList)
70 keep (list) ← [True] * n //list of TRUE Booleans of size n
71 FOR i IN RANGE 0 TO n
72    FOR j IN RANGE (i + 1) TO n
73       x1, y1 = centList[i][0], centList[i][1]
74       x2, y2 = centList[j][0], centList[j][1]
75       centroidDist ← sqrt((x2 – x1)^2 + (y2 – y1)^2)
76       IF keep[j] == TRUE AND centroidDist < 7
78          keep[j] ← FALSE
79
80 //Step 4: Remove overlapping hexagons
81 FOR i IN RANGE 0 TO length(hexList)
82    IF keep[i] == TRUE
83       APPEND hexList[i] TO hexagonsClean //can be used to draw hexagons
84       APPEND centList[i] TO centroidsClean
85
86 RETURN centroidsClean
```

**Table S3**. Pseudocode of the algorithm's helper function, find_neighbours()

```
    ALGORITHM: SEM pattern recognition & analysis
    HELPER FUNCTION: FIND_NEIGHBORS()
    Inputs
       centroids (list)... coordinates of centers of every valid contour
       est (float)... estimate for hexagon size
    Outputs
       linePointsClean (list)...start and endpoint of every line drawn
    ---------------------------------------------------------------------------------
87  linePoints(list) ← []
88  linePointsClean (list) ← []
89
90  FOR i IN RANGE 0 TO length(centroids)
91     distList (list) ← []
92     //Step 1: Find distance from centroids[i] with every other centroid
93     FOR j IN RANGE 0 TO length(centroids)
94        IF i != j
95           dist ← Euclidian distance from centroids[i] to centroids[j]
96           APPEND dist TO distList
97
98     //Step 2: Find 6 nearest neighbors
99     //can use argSort(distList) to find these
100    sixNearIndices (list) ← indices of 6 nearest neighbors
101    sixNearDist (list) ← distances of 6 nearest neighbors
102
103    //Step 3: Determine whether a line is "valid"
104    avgDist (float) ← sum(sixNearDist) / 6
105    stDev (float) ← standardDeviation(sixNearDist)
106    FOR j IN RANGE 0 TO 5
107       IF dist[j] <= (est + est /2) AND (dist[j] <= avgDist OR stDev <= 2)
108          APPEND [centroids[i], centroids[j]] TO linePoints
109
110 //Step 4: Remove duplicates
111 FOR i IN RANGE 0 TO length(linePoints)
112    IF linePoints[i] != any other item in linePoints
113       APPEND linePoints[i] TO linePointsClean
114
115 Return linePointsClean
```